\documentclass{aa}  
\usepackage{graphicx}
\usepackage{txfonts}
\usepackage{soul}

\usepackage[]{hyperref}
\hypersetup{
    colorlinks=true,
    linkcolor=blue,        
    urlcolor=blue,
    citecolor=blue,   }
    
\usepackage{xcolor}

\begin{document} 

   \title{Transformer neural networks for closed-loop adaptive optics using non-modulated pyramid wavefront sensors}

    \authorrunning{Weinberger et al}

   \author{Camilo Weinberger,
          \inst{1}
          Jorge Tapia,
          \inst{1}
          Benoit Neichel
          \inst{2}
          \and
          Esteban Vera\inst{1}
          }

   \institute{School of Electrical Engineering, Pontificia Universidad Cat\'olica de Valpara\'iso, Valpara\'iso, Chile\\
              \email{esteban.vera@pucv.cl}
         \and
            Aix Marseille Univ, CNRS, CNES, LAM, Marseille, France        
             }
   \date{Received December 1, 2023}

   \abstract
   {The Pyramid Wavefront Sensor (PyWFS) provides with the needed sensitivity for demanding future adaptive optics (AO) instruments. However, the PyWFS is highly nonlinear and requires the use of beam modulation to successfully close an AO loop under varying atmospheric turbulence conditions, at the expense of a loss in sensitivity.} 
   {This work aims to train, analyse, and compare the use of deep neural networks (NNs) as non-linear estimators for the non-modulated PyWFS, identifying the most suitable NN architecture for reliable closed-loop AO.}
   {We develop a novel training strategy for NNs that seeks to accommodate for changes in residual statistics between open and closed-loop, plus the addition of noise for robustness purposes. Through simulations, we test and compare several deep NNs, from classical to new convolutional neural networks (CNNs), plus a state-of-the-art transformer neural network \textcolor{black}{(TNN, Global Context Visual Transformer, GCViT)}, first in open-loop and then in closed-loop. By identifying and properly retraining the most adequate deep neural net, we test its simulated performance first in open-loop and then for closing an AO loop at a variety of noise and turbulence conditions. We finally test the trained NN ability to close a real AO loop for an optical bench.
   }
   {
   Using open-loop simulated data, we observe that a TNN (GCViT) largely surpasses any CNN in estimation accuracy in a wide range of turbulence conditions. Moreover, the TNN performs better in simulated closed-loop than CNNs, avoiding estimation issues at the pupil borders. When closing the loop at strong turbulence and low noise, the TNN \textcolor{black}{using non-modulated PyWFS data} is able to close the loop \textcolor{black}{similar} to a PyWFS with \textcolor{black}{$12\lambda/D$} of modulation. When raising the noise only the TNN is able to close the loop, while the standard linear reconstructor would fail, even when introducing modulation. Using the GCViT, we close a real AO loop in the optical bench achieving a Strehl ratio between 0.28 and 0.77 for turbulence conditions ranging from 6cm to 20cm, respectively.
   }
   {
   Through a variety of simulated and experimental results, we demonstrate that a Transformer Neural Network is the most suitable architecture to extend the dynamic range without sacrificing sensitivity for a non-modulated PyWFS. It opens the path for using non-modulated Pyramid WFSs under an unprecedented range of atmospheric and noise conditions.
}
   
   \maketitle
%

\section{Introduction}
The world will see the arrival of Extremely Large Telescopes (ELTs)—with primary mirrors larger than 25 meters of diameter—within the next ten years. However, atmospheric turbulence affects light propagation, acting as a dynamic phase mask that introduces aberrations to the optical path, finally diminishing the ability of current large telescopes, as well as future ELTs, to properly focus light at the diffraction limit. Adaptive Optics (AO, \citealp{roddier1999adaptive}) can assist modern telescopes to overcome atmospheric turbulence, firstly by measuring/estimating the wavefront fluctuations coming from reference sources, and then optically compensating for the aberrations before reaching the science instruments. Over the past 25-years, AO has revolutionized astronomy by providing the highest achievable image quality for ground-based observatories, becoming a fundamental component in the upcoming ELTs from first light \citep{AOELTreview}.

Wavefront sensors (WFSs) are the core of modern AO systems. Basically, a WFS needs to perform measurements, quickly enough (often within a millisecond), to infer the dynamic phase aberrations present in the wavefront passing through the atmosphere on its way to the telescope. Thus, an AO loop can use a deformable mirror to compensate for the atmospheric turbulence in real-time.
Within the known WFSs, the Pyramid Wavefront Sensor (PyWFS, \citealp{Ragazzoni1996}) exhibits relevant performance advantages--e.g. high sensitivity, large spatial frequencies and less noise propagation \citep{fauvarque2016general,chambnoise}--reasons why it has been successfully implemented in current large telescopes (LBT, \citealp{esposito2010first}, Subaru, \citealp{guyon2020validating}, and Keck, \citealp{mawet2022fiber}) and it is being considered for the next generation ELT instruments (HARMONI, \citealp{neichel2022harmoni} and MICADO, \citealp{clenet2022micado}). 

The PyWFS places the apex of a pyramidal 4-sided prism at the PSF plane (\textcolor{black}{focal} plane) of the incoming wavefront, finally re-imaging 4 different version of the pupil projected onto a detector array. Then, the slopes of the wavefront can be easily estimated from the measured image, although the PyWFS finally offers a very limited dynamic range where its response is still linear \citep{verinaud2004nature,Burvall:06}, despite its superb sensitivity.
In practical scenarios, and as suggested in the seminal work done by \citet{Ragazzoni1996}, the inherent non-linearity of the PyWFS can be counteracted by circularly modulating the incoming beam over the apex of the pyramid, which homogenize the illumination of the four sides of the prism. Although the linearity is improved, beam modulation comes with a detrimental effect on the sensitivity, plus the need for additional fast and expensive optomechanical elements. The non-linearity of the PyWFS becomes evident when using linear matrix-based reconstruction models \citep{Korkiakoski:07}. Despite  there are several efforts in the literature to use non-linear least-squares methods--which are often iterative--such as in \citet{Frazin:18,shatokhina2020review,hutterer2023mathematical}, there are also alternative approaches that exploit the compensation of the optical gains (OG) based on the turbulence statistics \citep{deo2019telescope,chambog}. Nevertheless, OG compensation is actually only a first order approximation of the non-linearities \citep{deo2019telescope}, being particularly hard to handle for the non-modulated PyWFS case. 

Nowadays, neural networks (NNs), and more particularly deep neural networks\textcolor{black}{–-that make use of a larger amount of hidden layers and intricate interconnections}, are a great asset to solve a variety of hard, non-linear problems in imaging such as detection, classification and inference \citep{lecun2015deep}. This is no different for WFSs, where in \citet{Nishizaki:19} they demonstrated that any imaging system can be turned into an image-based WFS by appropriately training a convolutional neural network (CNN) to infer the incoming wavefront Zernike modal approximation. Since then, deep learning has been applied to improve the estimation performance of focal-plane WFSs \citep{orban}, Shack-Hartmann WFSs \citep{DuBose:20}, phase diversity-based WFSs \citep{andersen2020image}, Lyot-based WFSs \citep{Allan:20}, and also the PyWFS \citep{Landman:20}, where they use a CNN jointly with the linear estimator to improve the PyWFS linearity. 
Recently, \citet{wong2023nonlinear} demonstrated the ability of a three-layer fully connected neural network to estimate low-order modes from a PyWFS with and without modulation, while \citet{archinuk2023mitigating} used a simple CNN to estimate the first 400 modes from the non-modulated PyWFS. Since AO is a two stage process--wavefront sensing and wavefront control--we can also use deep NNs not only to improve the \textcolor{black}{accuracy of the wavefront estimation} of the WFS, but also the performance of the closed-loop AO system, as proposed in \citet{Nousiainen:21} and \citet{Pou:22}, where they use reinforcement learning.

Most of the deep learning WFSs have been adapting conventional deep neural nets originally developed for computer vision applications, such as Xception \citep{Chollet2017}, VGG--Net \citep{Simonyan15}, or ResNet \citep{He2015DeepRL}, while in \citet{Vera2021} the authors crafted an original deep neural net (WFNET) for image-based WFSs. Nevertheless, there is a new generation of CNNs--ConvNeXt and ConvNeXt v2 \citep{convnext2020,convnext2}--that are delivering impressive performance in classification tasks. Moreover, there is a new class of deep neural nets, called Transformers Neural Networks (TNN), that are overcoming the performance of classical CNNs for a variety of vision applications, being the Visual Transformer ViT \citep{VITpaper}, and the Global Context Visual Transformer GCViT \citep{GCViTpaper}, the current state-of-the-art. Instead of convolutions, TNNs such as the GCViT can find correlations between image patches and their influence on the target outputs. Nevertheless, the main disadvantage of most modern CNNs and TNNs is that they often require massive amount of data for training, although they can be efficiently retrained without requiring to start from scratch.  

We believe that a good wavefront estimator/reconstructor will necessarily lead to a good AO performance. Therefore, this work aims to study and analyse the use of deep neural nets as a non-linear estimator for the non-modulated PyWFS, enabling reliable open-loop and closed-loop adaptive optics performance. For that, we will test and compare the ability of several conventional and newest neural network architectures to handle the PyWFS non-linearity, developing appropriate training strategies to accommodate for changes in residual statistics between open and closed-loop. By choosing and properly training the most adequate deep neural net for the task, we believe we can demonstrate the ability to extend the dynamic range of the non-modulated PyWFS without sacrificing sensitivity, enabling closed-loop operations at a variety of noise and turbulence conditions without requiring beam modulation anymore.  
   
\section{Methodology}
In this section we describe the methods used for: (1) the simulation of the PyWFS forward model under a variety of turbulence and noise conditions, (2) the training of the deep neural networks using open-loop data, (3) the retraining strategy for closed-loop scenarios, (4) the metrics to quantifying the performance of the different NN estimators, and (5) the configuration of the experimental setup.

\subsection{Simulation framework}
We use the OOMAO toolbox \citep{conan2014object}, written in Matlab, to generate the incoming aberrated wavefronts and simulate the propagation through the PyWFS up to the detector plane. The phase map dataset is generated with a spatial resolution of \textcolor{black}{$268\times268$} pixels for a $1.5$m aperture telescope working at $\lambda = 550 nm$ with an $r_0$ distribution ranging from $1$cm to $20$cm (distributed in discrete steps of $2$cm after $2$cm), leading to an effective $D/r_0$ range between $7.5$ and $150$. \textcolor{black}{It is important to note that as D/r0 increases, the turbulence becomes stronger}. For every phase map, we retrieve the first 209 Zernike coefficients, ignoring piston, which are considered as the ground-truth for estimation purposes. After propagating each phase map for an incoming Magnitude 0 light source in the V-band \textcolor{black}{using OOMAO}, we store the intensity image $I$ projected on the \textcolor{black}{WFS} detector of size \textcolor{black}{$268\times268$} pixels with an exposure \textcolor{black}{time} of 1 sec without \textcolor{black}{considering} noise. \textcolor{black}{The diameter of each subpupil at the simulated WFS detector span 68 pixels}. In total, the open-loop phase map dataset is comprised of 210,000 uncorrelated samples (phase map, Zernikes and PyWFS image) obtained from random turbulence realizations given the selected $D/r_0$ level. 

Additionally, we also simulated phase map sequences (using two layers moving at 5 and 10m/s sampled at 250Hz) of 1,000 samples for different turbulence strengths and several levels of modulation for the PyWFS, at $0\lambda/D$ (non-modulated), $3\lambda/D$, $5\lambda/D$, and $12\lambda/D$. These sequences are aimed for closed-loop testing.

\subsection{Magnitude and noise}
We added noise to the intensity image $I$ on demand as required either by the training or testing stage. By fixing the telescope size to $1.5$m, we first scale the image to a proportional photon flux depending on the magnitude of the star ($\text{Mag}$) and the exposure time ($T_e$) such that  
\begin{equation}
\text{I}_{ref} = \text{I} \cdot 10^{-0.4\cdot\text{Mag}} \cdot T_{e}.
\end{equation}
Then, we apply Poisson noise to the scaled measurement $\text{I}_{ref}$ leading to the noisy image $\text{I}_{noise}$, enabling the calculation of the effective \textit{signal--to--noise ratio} (SNR) defined as
    \begin{equation}
        \text{SNR} = \frac{\sigma(\text{I}_{ref})}{\sigma(\text{I}_{ref}-\text{I}_{noise})}.
    \end{equation}
    
\begin{figure}[h!]
        \centering
        \includegraphics[width=\hsize]{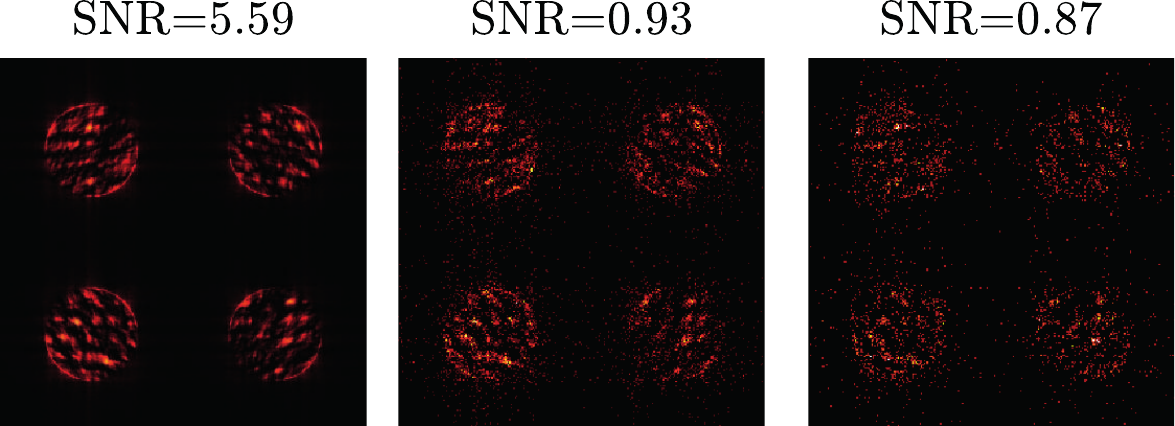}
        \caption{Non-modulated PyWFS measurement for different star magnitudes at a fixed exposure leading to different effective SNR levels.}
        \label{fig:nPhotons}
    \end{figure}

Figure~\ref{fig:nPhotons} displays the example of three images measured by the PyWFS for different star magnitudes at a fixed exposure time, with the respective associated SNR. We can observe a clear degradation of the image as the photon flux diminishes. Before being used with the linear or NN estimators, the noisy PyWFS measurement is always normalized. Throughout the article, we only refer to the SNR level of the measurements when dealing with noise.
    
\subsection{Neural network training}
We selected four deep neural network architectures as non-linear estimators for the non-modulated PyWFS. Xception~\citep{Nishizaki:19}, WFNet~\citep{Vera2021}, ConvNext~\citep{convnext2020}, and GCVit~\citep{GCViTpaper}, are implemented in PyTorch and adapted to perform regression at the last layer to provide simultaneous estimations for the first $209$ Zernike modes (without piston).
    
\textcolor{black}{Since all implemented NNs accept the same input image size from the PyWFS and deliver estimates of the Zernike coefficients with same number of coefficients as well, then} all NNs are trained and tested with the exact same portions of the simulated dataset, which is divided into $75\%$ for training and $25\%$ for testing. All training sessions were performed in PyTorch by using 8 NVIDIA Quadro RTX5000 GPUs. From several preliminary tests on the NNs, we realized that the choice of a proper range of values for the learning rate and a suitable loss function heavily depend on the strength and amount of Zernikes modes to be estimated, which are also related to the turbulence strength. \textcolor{black}{A loss function is an error metric that is computed between the estimated output values of the NN (in this case a vector of Zernike coefficients $\hat{z}$) and the vector of ground-truth values ($z$) used for training. The most common loss functions are the mean square error (MSE) and the mean absolute error (MAE), defined as}

\textcolor{black}{
\begin{equation}
\text{MSE}=\frac{1}{N}\sum_N{(z-\hat{z})^2} \qquad \text{and} \qquad \text{MAE}=\frac{1}{N}\sum_N{|z-\hat{z}|},
\end{equation}}

\noindent \textcolor{black}{where $N$ is the number of coefficients. On the other hand, the learning rate is the weight given to the calculated loss function that is back-propagated to update the NN hidden parameters at every training iteration}. For instance, mixing a large learning rate (larger than $10^{-5}$) with \textcolor{black}{the MAE} loss function allows a correct training of low-order, high-amplitude Zernikes modes.
In contrast, mixing a small learning rate ($\approx10^{-6}$ or less) with \textcolor{black}{the MSE} improves the training of high-order, low-amplitude Zernikes modes. 

\begin{figure}[!h]
        \centering
        \includegraphics[width=\hsize]{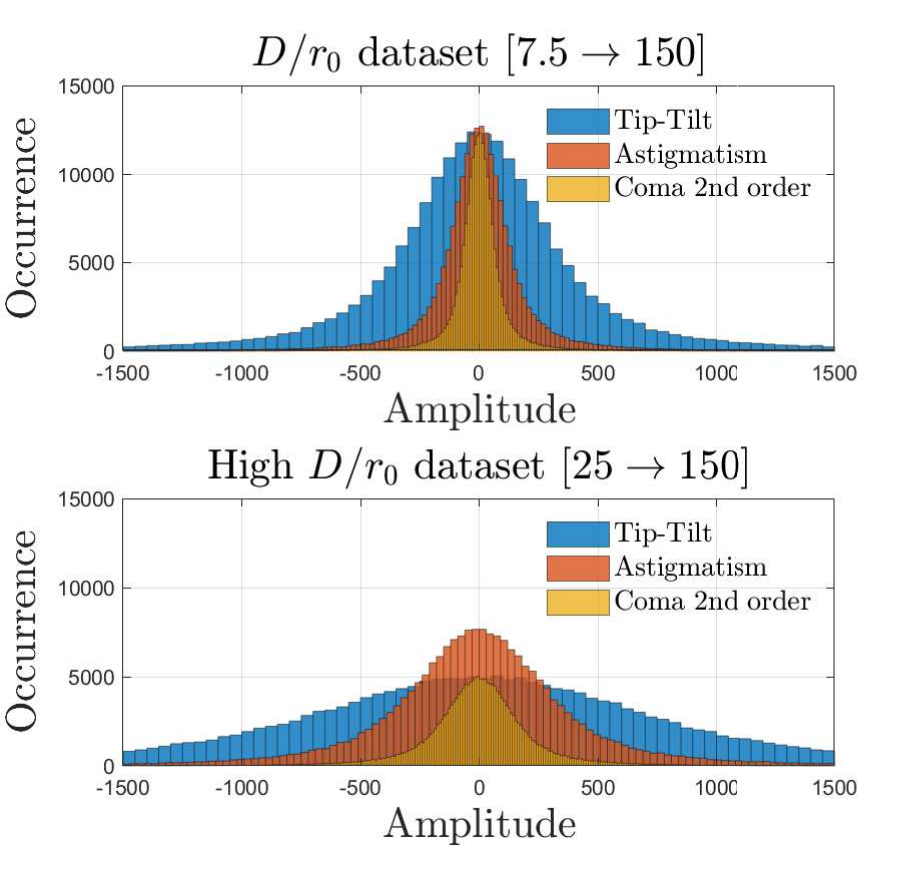}
        \caption{Amplitude distribution for selected Zernike modes for two turbulence training regimes. Full range of $7.5 > D/r_0 > 150$ on the top and high range of $25 > D/r_0 > 150$ on the bottom.}
        \label{fig:histogram}
    \end{figure}

Therefore, we generated a two-step training strategy for open-loop wavefronts, creating two training datasets with different distributions for the Zernike modes, as seen in Figure~\ref{fig:histogram}.
The first training uses a limited dataset with the range between $D/r_0=25$ and $150$, a starting learning rate of $10^{-5}$, and MAE as the loss, which emphasizes a boost in the linearity response of the non-modulated PyWFS. 
Then, in the second stage we retrain the NNs using the whole dataset from $D/r_0=7.5$ to $150$, a starting learning rate of $10^{-6}$, and MSE as the loss, which plays a significant role in preserving the sensitivity of the PyWFS.

\subsection{Closed-loop training}
\label{section:closed_loop}
\begin{figure}[!h]
      \centering
      \includegraphics[width=\hsize]{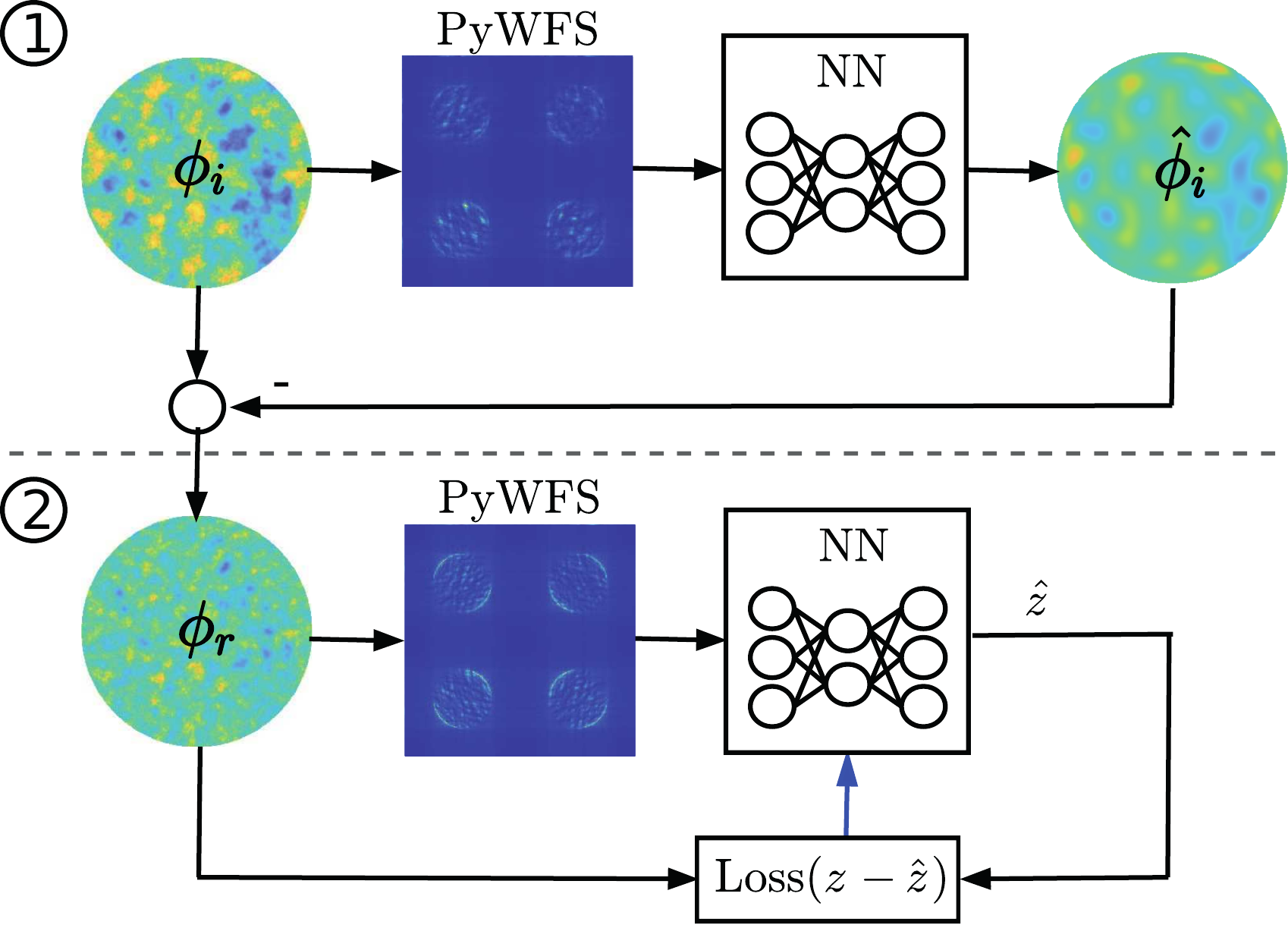}
      \caption{Neural network closed-loop training strategy. \textcolor{black}{The first stage uses the estimation by the NN from open loop data to compute a residual phase which is used as the input for the second stage. The Zernike coefficients of the residual are used as the ground-truth to compute the loss function used to retrain the NN.}}
      \label{fig:closed-loop-scheme}
  \end{figure}

After training the NNs to properly estimate the output from a non-modulated pyramid for a variety of turbulence conditions in open loop, we devise a retraining strategy to prepare for the statistics of the residuals in closed-loop. We propose a two-step approach as depicted in Figure~\ref{fig:closed-loop-scheme}. In the first step, we input a phase map from the dataset to the PyWFS and estimate the Zernike coefficients from the chosen NN architecture. Then, we reconstruct the estimated phase out from the Zernike coefficients and compute the phase residual by plain subtraction. In the second step, we input the phase residual to the PyWFS and estimate a new set of Zernike coefficients using the NN. Now we calculate the loss function MAE for the estimated residual coefficients, and update the parameters of the NN with a learning rate of $10^{-6}$.
  
This novel training approach for closed-loop measurements allows to use the simulated wavefronts in the dataset independently, without worrying if they are correlated or not in time. \textcolor{black}{The proposed scheme serves as an effective data augmentation approach}, which leads to a higher diversity in the statistics provided for the NN models. 

In our initial training using closed-loop data we consider an ideal noiseless case. Nonetheless, when we have to prepare a NN for  more realistic scenario, we retrain the NN by randomly selecting a SNR level for the PyWFS measurement (between SNR=0.7 and SNR=7), which tends to improve the robustness of the trained NNs \citep{bishop1995neural}.

\subsection{Performance metrics}
For comparing the open-loop \textcolor{black}{wavefront estimation accuracy} between the different NNs in simulations, we use the Root Mean Square Error (RMSE) of the predicted Zernike coefficients as follows, 

    \begin{equation}
    \text{RMSE} = \sqrt{\frac{1}{N}\sum{(\textcolor{black}{z} - \textcolor{black}{\hat{z}})^2}},
    \end{equation}

\noindent where \textcolor{black}{$z$} corresponds to the ground truth Zernike coefficients extracted from the incoming wavefront $\phi{i}$, \textcolor{black}{$\hat{z}$} are the Zernike coefficients estimated by the NN, and $N$ is the number of Zernike coefficients.

When we switch to closed-loop in simulations, we analyze the standard deviation of the phase map residual  $\sigma_{\phi}$. The residual is computed by the difference between the incoming phase $\phi_i$ and the corresponding update given by the last reconstruction from the last estimated Zernike coefficients.

Also, for the experimental validation, we  analyze the system performance by computing the Strehl ratio (SR) given by the ratio between the maximum value of the reconstructed PSF and the maximum value of the equivalent diffraction limited PSF. 
  
\subsection{Experimental AO bench}
\begin{figure}[h!]
        \centering
                \includegraphics[width=\hsize]{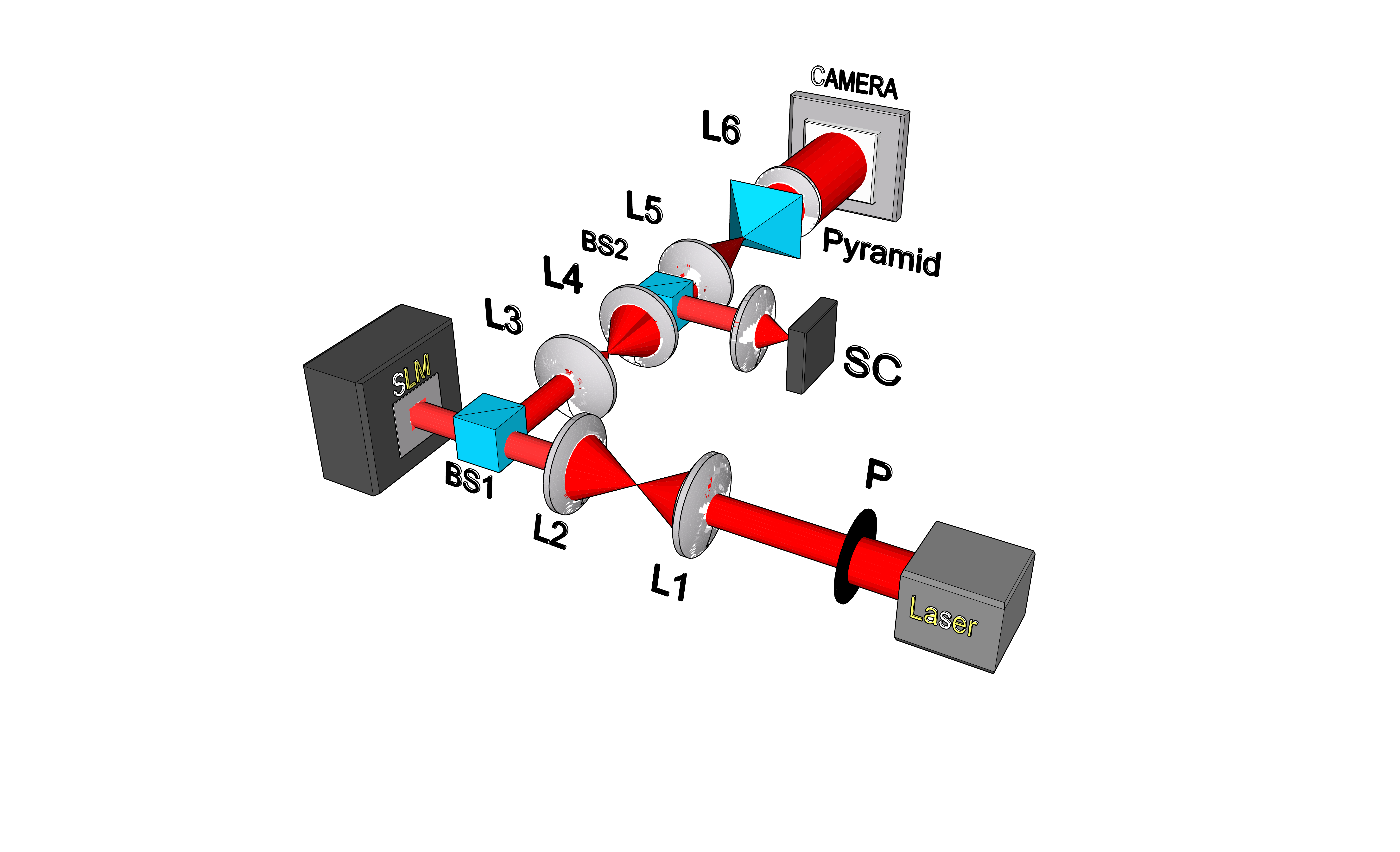}
        \caption{Schematic of the experimental AO setup using PULPOS to test the PyWFS in open- and closed-loop.}
        \label{setup}
    \end{figure}

We use the PULPOS AO bench~\citep{Tapia2022} to validate the performance of the PyWFS + GCVIT in closed-loop. The particular branch of PULPOS used to close the AO loop with a PyWFS is shown in Fig.~\ref{setup}. We use a $\lambda = 635 nm$ fiber-coupled laser source (Thorlabs S1FC635) attached to an air-spaced doublet collimator (Thorlabs F810APC-635) and a beam expander (Thorlabs GBE02-A). After a 5 mm diameter aperture stop (P), the beam passes through a 4f-system with 1X magnification (L1 and L2) before reaching the reflective high-speed spatial light modulator (SLM, Meadowlark HSP1920-488-800-HSP8, \textcolor{black}{$1920\times1152$ pixels, $9.2\mu m$ pixel size}), where phase maps \textcolor{black}{of $560\times560$ pixels are projected} to emulate the desired turbulence, \textcolor{black} {matching the pupil size relayed at the SLM}. Then, a beamsplitter (BS1) redirects the aberrated wavefront through a 0.75X magnification 4f-system (L3 and L4), reaching the L5 lens (400 mm) that focus the beam on the \textcolor{black}{PSF} plane where the apex of a zeonex pyramid is located. Then, the L6 lens (200 mm) collimates the four sub-pupils emerging from the pyramid, projected onto a high-speed CMOS camera (Emergent Vision HR-500-S-M, 9$\mu m$ pixel size, 1586 fps, \textcolor{black}{$812\times620$ pixels}).\textcolor{black}{The WFS images are cropped at $620\times620$ pixels, where each of the subpixels span a diameter of 110 pixels. These images are resized to match the subpixel diameter of the pupils used in the simulations before entering the NN estimation. In parallel, we record the PSF that is imaged by a $125mm$ lens onto the science camera (SC, Emergent Vision HR-500-S-M, $9\mu$m pixel size), where we extract the Strehl ratio}. The SLM and cameras are controlled by a desktop computer loaded with a RTX4000 GPU. 

\section{Results}
In this section, we start by going through the results obtained in open- and closed-loop using simulations of the non-modulated PyWFS estimated with a variety of deep NN options. Then, these results drove the decision of selecting the best performing NN architecture, which is used in the final experimental validation using PULPOS in open- and closed-loop as well.

\subsection{Neural network comparison}
  \label{results:OL}
In our initial test results, we evaluated the performance of several deep NNs trained with identical parameters utilizing the entire dataset range ($r_0 = [1 \to 20]$ cm). The chosen NNs are three CNNs, Xception, WFNet, and ConvNext, plus one TNN, GCViT. In particular, we trained the lightweight version of the GCViT, which is the GCViT--xxtiny.

In Table \ref{table:openloop_result}, we present a summary of the number of parameters, the estimation speed--for the same training computer machine using a single GPU card--and the estimation performance (average fitting error per mode) from noiseless measurements of every tested NN architecture. \textcolor{black}{The number of parameters refers to the number of interconnections (weights) inside each NN, which depends on the number of layers and neurons of each architecture and has great correlation to the size of the NN stored in the GPU memory. However, every architecture has its own intricacies and choice for the number of hidden layers and the inner operators such as linear convolutions and neuron non-linearities that finally affect in different manner the speed of calculus of each NN.}

    \begin{table}[h!]
        \caption{\textcolor{black}{Comparison of deep neural networks used for WFS in terms of the} number of parameters \textcolor{black}{(\# Params)}, inference speed \textcolor{black}{(Speed)}, and estimation error \textcolor{black}{(Error)}. The best result is highlighted within each column.}             
        \label{table:openloop_result}      
        \centering                          
        \begin{tabular}{c c c c}        
            \hline\hline                 
            NN & \# Params ($10^6$) & Speed (ms) & Error (nm) \\
            \hline                        
               Xception     & $22$    & $1$                &  $32.5 \pm 7.0$ \\      
               WFNet        & $152$   & $\textbf{0.37}$     &  $286.8 \pm 28.3$ \\
               ConvNext     & $88$    & $1.13$               &  $102.5 \pm 16.3$ \\
               GCVit--xxtiny & $\textbf{12}$   & $1.56$      &  $\textbf{25.0} \pm {7.2}$ \\
               \hline                                   
        \end{tabular}
    \end{table}

From the results, it is evident that the GCViT achieves the best average performance, meaning that it is able to provide with good estimates in the whole turbulence range. A slightly worse performance is surprisingly achieved by the Xception, despite being an older CNN in contrast with the newer ConvNext. The worst performance was achieved by the WFNet, maybe related to the fact that it was originally developed for undersampled image-based WFS. Interestingly, we can also observe that the error was somehow proportional to the number of parameters of the NNs. Nevertheless, the GCViT seems to be the slowest, despite having fewer parameters. One possible explanation for this is the fact of being a way more intricate and complicated architecture.
    
\begin{figure}[h!]
        \centering 
         \includegraphics[width=\hsize]{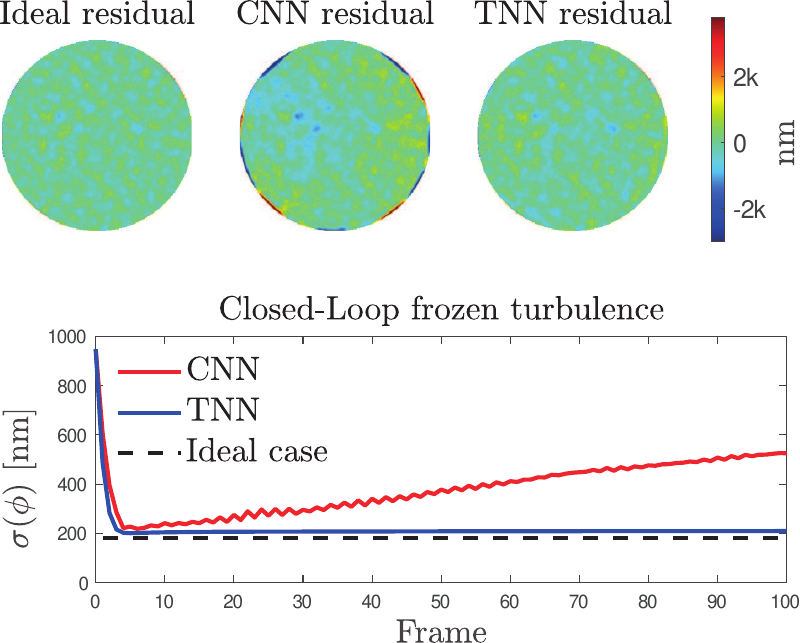}
        \caption{Closed loop performance in simulations using a constant input phase with two neural networks, a CNN (Xception) and a TNN (GCViT). Top: last frame of the closed-loop residual phase map. Bottom: Evolution of the residual standard deviation, comparing the CNN, TNN and the optimal estimation of 209 Zernike modes.}
        \label{fig:frozenT}
\end{figure}

To select the most promising NN candidate for real AO applications, we test the best two performing NNs in closed-loop, the Xception as the CNN candidate and the GCViT as the TNN one. We performed the close loop test under a ``frozen'' turbulence condition--static phase map at the input--and no noise (SNR$=\infty$) to analyze their estimation and compensation behavior in ideal conditions. Closed loop results for a turbulence of $D/r_0=75$ are depicted in Figure \ref{fig:frozenT}, where we can observe that both NNs quickly reduce the residual within a few frames. Nevertheless, at some point the CNN tends to diverge as the TNN approaches the ideal residual value for the estimation of 209 Zernike modes. This problem for the the CNN in closed loop is clearly seen after inspecting the residual phase map, where significant aberrations start to appear at the borders, most likely created by a deficient estimation of the high order modes by the CNN. 

We believe this behavior of the CNN must be due to the convolutional nature of the CNN, which may complicate the handling of the phase at the sharp pupil borders. On the other hand, the TNN can identify the portions within the image that are informative, which may explain its superior performance, delivering a spatially homogeneous residual phase map. As a side note, please be aware that it is impossible to close the loop using a linear estimator for the non-modulated PyWFS at this turbulence level.
    
\subsection{Noise response}
\begin{figure}[h!]
        \centering
        \includegraphics[width=1\hsize]{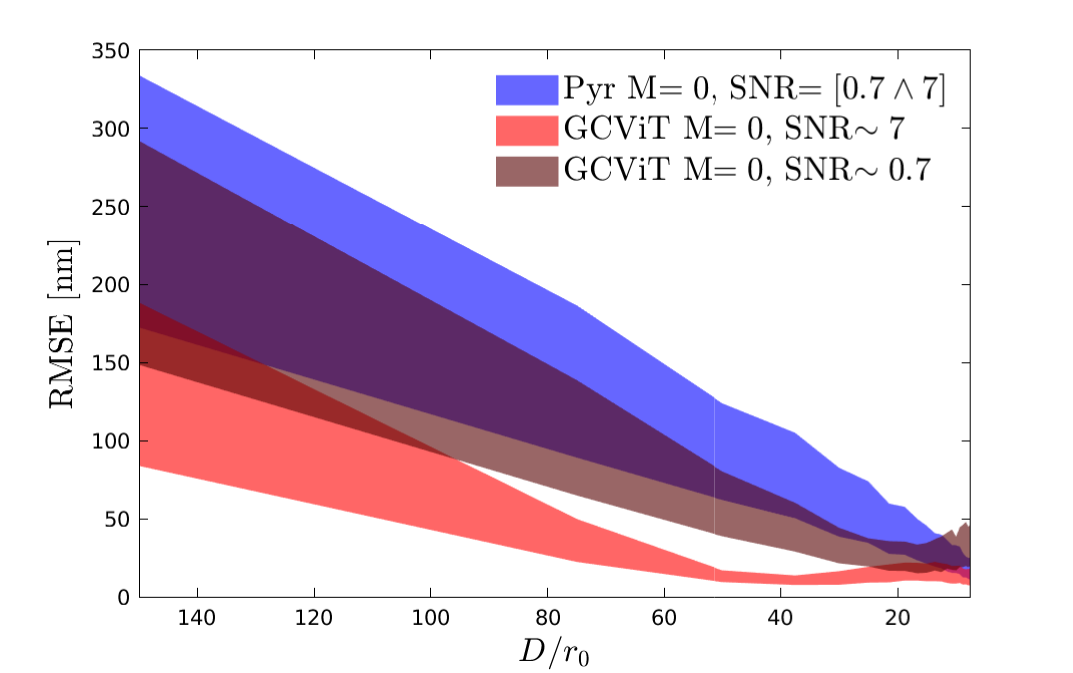}
        \caption{Open loop performance comparison in simlations for the linear least-squares estimation and the GCViT estimation for a non-modulated PyWFS at different SNR.} 
        \label{fig:openloop_comp}
    \end{figure}

Once we decided that the GCViT is the most suitable NN architecture for closing the loop with a non-modulated PywFS, we retrained the GCViT with different levels of photon noise equivalent to a range of SNR between $0.7$ and $7$, randomly applied to the measurements. We compare the GCViT estimation with the linear estimation of the non-modulated PyWFS for a turbulence range between $D/r_0 = 7.5$ and $D/r_0 = 150$ in open-loop. Performance results for measurements taken with two noise levels--low noise (SNR=7) and high noise (SNR=0.7)--are presented in Fig.~\ref{fig:openloop_comp}. The first observation is that the linear estimation for the non-modulated PyWFS is extremely immune to noise, given its high sensitivity, so the two plots are merged into one. 

For a high SNR, the GCViT is vastly superior to the linear estimation in the whole turbulence range, clearly improving the linearity of the PyWFS response, particularly within the $D/r_0 = 20-60$ range. Although the GCViT estimations are still better than the traditional linear estimation for the PyWFS under low SNR conditions for most of the turbulence range, the linearity advantages are not as high as in the high SNR case. Nevertheless, the estimation for the GCViT can become slightly worse than the linear estimation for the PyWFS for very weak turbulence at $D/r_0=7.5$.

\subsection{Closed-loop performance}
\begin{figure*}[!h]
        \centering
         \resizebox{\textwidth}{!}
            {\includegraphics[bb= 00 0 1260 640,clip]{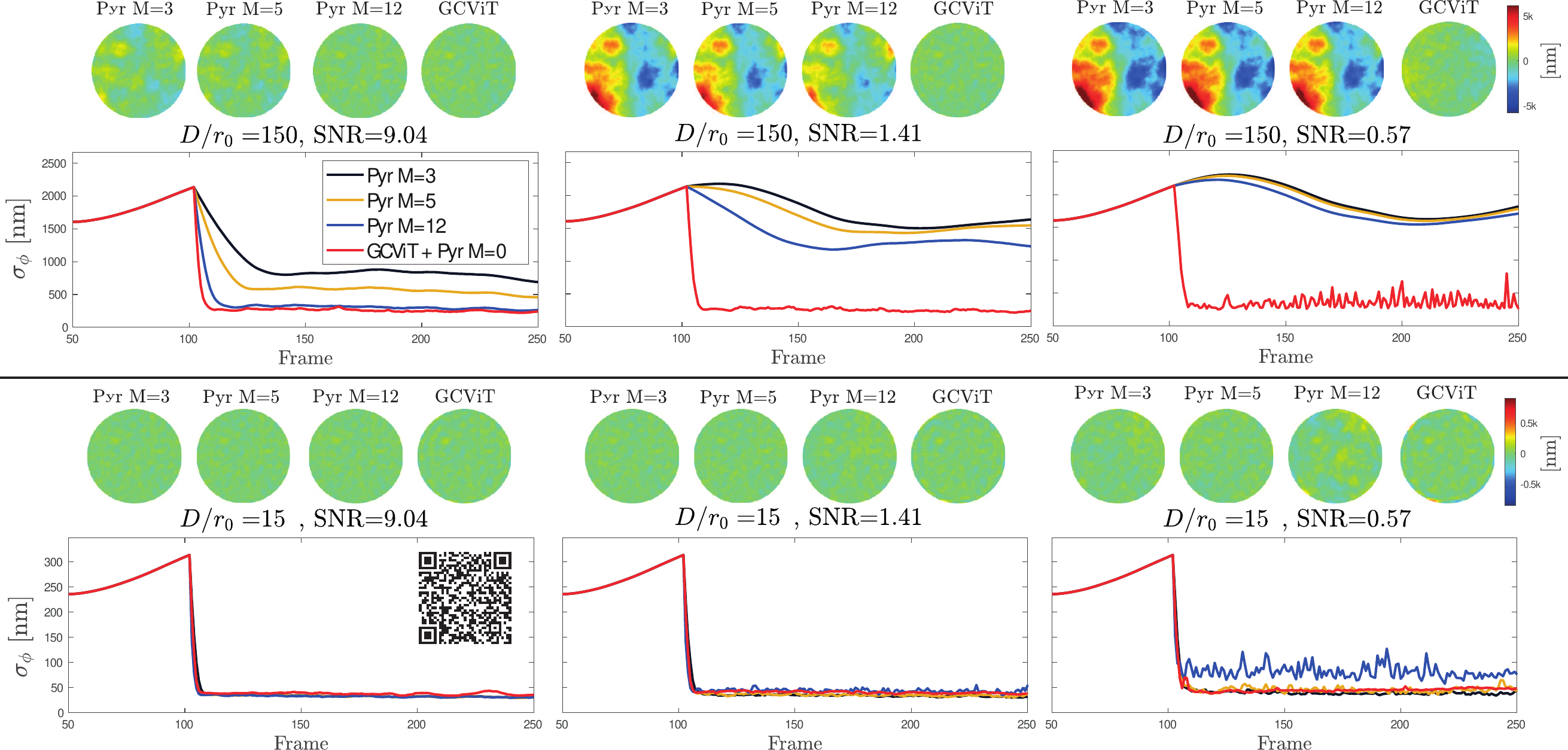}}
            
        \caption{\href{https://www.youtube.com/playlist?list=PLdasfH9Ad7BBxRzNY80DzAHecytL2XZoc}{(Movie online)} Closed loop residual phase evolution using simulated data at $D/r_0=150$ \textcolor{black}{(top column) and $D/r_0=15$ (bottom row)}, comparing the linear least-squares estimation for the PyWFS at different modulation levels (M={3,5 and 12}$\lambda/D$) against the GCViT estimation for the non-modulated PyWFS (M=0). The AO loop is closed at frame $100$ using a proportional controller with gain $k=0.5$.  From left to right: results for different SNR levels. At the top of each plot: residual phase for the different estimation methods at frame $250$.}
        \label{figure:CL}
    \end{figure*}

We test the performance of the GCViT trained with noise when closing an AO loop using the non-modulated PyWFS. We compare the closed-loop performance against the PyWFS working at different modulations of $3\lambda/D$, $5\lambda/D$, and $12\lambda/D$, using the traditional linear least-squares estimator. We choose to close the AO loop \textcolor{black}{at weak turbulence conditions of $D/r_0 = 15$ as well as at the worst} trained turbulence conditions of $D/r_0=150$, where it is impossible to close the loop with the non-modulated PyWFS using linear estimation. In Figure \ref{figure:CL} we present the results when closing the AO loop for three different SNR levels: $9.04$, $1.41$, and $0.57$. Note that the lowest SNR level is beyond the training regime used for the GCViT.   

The AO loop is closed at the $100^{th}$ frame. \textcolor{black}{For the strong turbulence case (shown at the top row)} we can observe that for a high SNR, all the WFSs are able to \textcolor{black}{close the loop and} reach a stable \textcolor{black}{but different residual} levels, being the \textcolor{black}{GCViT and the} PyWFS at $12\lambda/D$ using linear estimation the best \textcolor{black}{performing, indicating that} they may share an equivalent linear response at that turbulence regime, as corroborated by the very similar residual shape for the very last frame. As the SNR decreases, only the GCViT is able to keep up a similar level of residual as obtained in high SNR conditions, while the PyWFS using linear estimation at different modulations is not successful in closing the loop. Only at the very lowest SNR level, the GCViT shows some slight difficulties for maintaining the expected residual levels achieved for higher SNRs, but without loosing the ability of closing the loop at all. By being able to close the loop at this extreme turbulence regime even at high noise levels, it seems that by using the non-modulated PyWFS measurements the GCViT is able to keep some of the inherent high sensitivity while still dramatically increasing the linearity. 
\textcolor{black}{The bottom row of Figure 7 reveals the results for a weak turbulence regime, where the GCViT shows a similar behavior, perhaps slightly worse in terms of residual, successfully closing the AO loop as all the modulated PyWFS versions for the high SNR scenario. However, as the SNR is decreased, we can easily note the loss of sensitivity for the PyWFS as modulation is increased at $12\lambda/D$, while the GCViT is able to maintain a stable residual in between of the PyWFS at $3\lambda/D$ and $5\lambda/D$ of modulation for the lowest SNR case.}

\subsection{Experimental validation}
\begin{figure}[h!]
        \centering
        \includegraphics[width=\hsize]{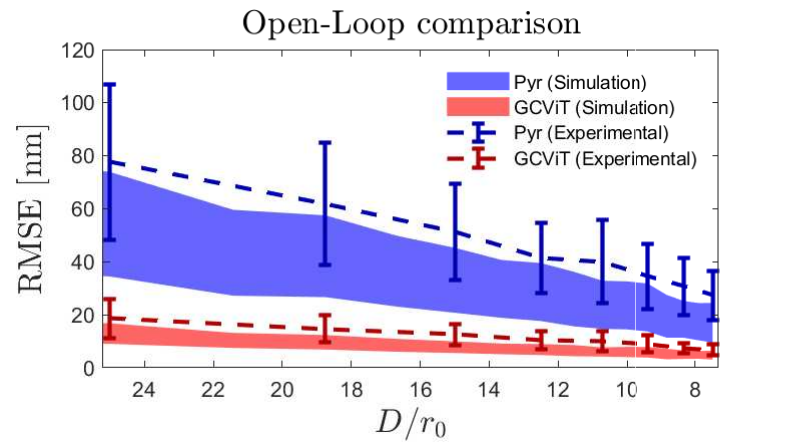}
        \caption{
        \textcolor{black}{Comparison of the wavefront estimation error} using least-squares estimation (Pyr) and the GCViT estimation for simulated and experimental open loop measurements from a non-modulated PyWFS.}
        \label{fig:simul-exper}
    \end{figure}

We use the PULPOS \citep{Tapia2022} AO bench to obtain measurements from the non-modulated PyWFS under controlled, arbitrary phase maps projected onto the SLM display. We start by calibrating the PyWFS and obtaining the interaction matrix by projecting the first pure 209 Zernike modes (without piston) in push and pull using an amplitude of $0.5\lambda$. As a first part of the experimental validation, we test the PyWFS using the same open-loop dataset used in the simulations, although only for $r_0= [6 \to 20]$ cm, which are the conditions found at the $1.5$m telescope at the Observatoire de Haute-Provence (OHP) that is currently running PAPYRUS \citep{papyrus}: an adaptive optics instrument based on a PyWFS. We present the performance results for the classical linear estimation method and the GCViT in Figure~\ref{fig:simul-exper}, comparing both the use of simulated measurements and the experimental PyWFS data obtained at PULPOS. Please note that the GCViT has solely be trained using simulated data.

We can notice from the plots in Fig.~\ref{fig:simul-exper} that the GCViT estimation using experimental data vastly outperforms the classical least-squares estimation, as predicted by the simulations. Despite some general offset in both cases, the experimental estimations follow the overall trend and standard deviation obtained when using the simulated measurements. The linearity offered by the GCViT estimations using the non-modulated PywFS are superior to what is being offered by the linear estimation methods. On top of that, we may even consider that the sensitivity offered by the GCViT is also better, considering the superior performance at low turbulence levels and that measurements are noisy since the CMOS camera being used for the PyWFS is not a scientific-grade camera.

\begin{figure}[h!]
        \centering
        \includegraphics[width=\hsize]{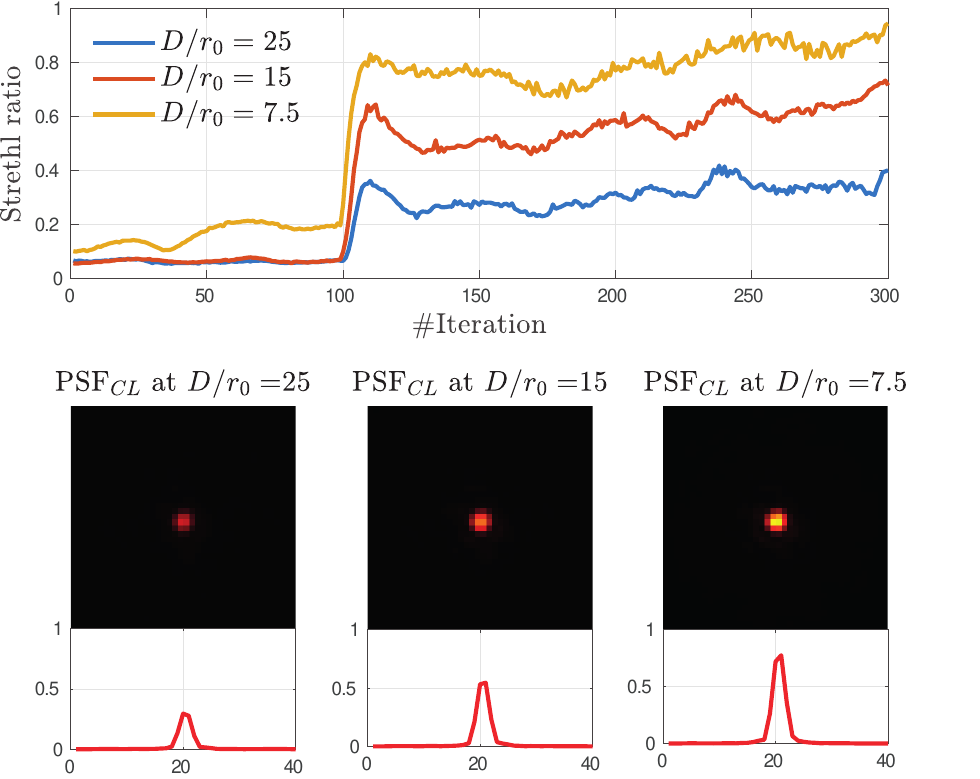}
        \caption{\href{https://www.youtube.com/watch?v=wSQBWTgDHTs&list=PLdasfH9Ad7BB58D2VabF1SzNqx-MHV85W}{(Movie online)} Experimental closed-loop performance at different turbulence conditions using GCViT with a non-modulated PyWFS. The AO loop is closed at frame $100$ (gain $k=0.3$). Top: Evolution of the strehl ratio for different turbulence conditions. Bottom: Integrated PSFs in closed-loop (PSF$_{CL}$) with their respective horizontal line profiles.}
        \label{fig:Experimental_CL}
    \end{figure}

As the second part of the experimental validation, we use the trained GCViT to close an AO loop using PULPOS with the non-modulated PyWFS. 
We estimate the 209 Zernike modes on-the-fly using a NVIDIA RTX4000 GPU. As a deformable mirror, we use the same SLM where we project the aberrated turbulence phase maps in open loop, now displaying the compensated phase maps given by the chosen control law applied to the estimations provided by the GCViT. Although the simulated phase map sequence used for closed-loop is sampled at $250$Hz, we close the AO loop in PULPOS at $10$Hz, as we are working on a realtime control upgrade. We choose to close the loop at three representative turbulence conditions with $r_0$ at $6$, $10$, and $20$ cms, equivalent to a $D/r_0$ of $25$, $15$, and  $7.5$ at OHP, respectively. Results displaying the evolution of the Strehl ratio--computed from the instantaneous PSFs captured by the scientific camera in PULPOS--for different turbulence conditions are shown at the top of Figure~\ref{fig:Experimental_CL}, where we close the loop at the $100^\text{th}$ frame. We can clearly observe that by using the GCViT we are able to close the AO loop in all situations, and that the loop is stable. As a side note, we can hardly close the loop at $D/r_0=7.5$ when using linear estimation with the non-modulated pyramid in the AO bench, explaining why these results are not even presented here.

We display at the bottom of Fig.~\ref{fig:Experimental_CL} the PSFs and respective central horizontal line profiles, integrated between frames $100$ and $300$, as obtained by the GCViT in closed-loop for different turbulence conditions, corresponding to an average SR of $0.28$, $0.56$, and $0.77$ for an $r_o$ of $6$cm, $10cm$, and $20$cm, respectively. 
   

\section{Conclusions}
In this work we presented a comparative analysis of using deep neural networks as non-linear estimators for the non-modulated PyWFS. We trained, tested and compared several conventional and state-of-the-art neural network architectures to handle the PyWFS non-linearity, where Convolutional Neural Networks have been the widest used architecture for WFS applications, so far. We developed a novel training strategy that combines the use of open and closed-loop data, as well as the addition of noise for robustness purposes. Through simulations, we have found that a state-of-the-art Transformer Neural Network, in this case the GCViT, is the most suitable NN architecture for closed-loop AO operation, avoiding systematic phase estimation problems at the pupil borders caused by the convolutional nature of CNNs. 

When testing in open-loop, the GCViT is able to dramatically extend the dynamic range of the non-modulated PyWFS in contrast with the traditional linear estimation methods at a variety of noise conditions, although there is a clear loss in performance at very high noise levels. 
However, when testing in a simulated AO scenario at the \textcolor{black}{worst} turbulence conditions, we have found that the GCViT is able to close a stable AO loop \textcolor{black}{similar} to a modulated PyWFS at \textcolor{black}{$12\lambda/D$} at a high SNR level. Moreover, the GCViT showed to be very robust to noise, since it was the only estimator able to reliably close the AO loop for mid to low SNRs \textcolor{black}{for strong turbulence conditions. When dealing with weak turbulence, the GCViT was able to close the loop for the whole range of SNR, with a performance similar to the PyWFS with $5\lambda/D$ of modulation}. These results were experimentally validated in the PULPOS AO bench, where the GCViT was able to consistently close the AO loop for turbulence ranging from $6$cm to $20$cm, achieving an integrated Strehl ratio at the scientific camera between $0.28$ and $0.77$, respectively. 

In conclusion, we demonstrated that a TNN such as the GCViT can be properly trained and become suitable as a non-linear estimator for a non-modulated PyWFS. By dramatically extending the dynamic range of the non-modulated PyWFS without sacrificing sensitivity, the GCViT can be used in real AO scenarios, being robust to noise and varying turbulence conditions, paving the way for further testing in real observing conditions, \textcolor{black}{which is out of the scope for this work. The fact that the proposed NN has been trained entirely offline is interesting. The presented experimental results showed that the trained TNN actually possesses a certain degree of flexibility to accommodate for statistical variations. Nevertheless, we believe that retraining using real data may be necessary in some cases, which will be validated through on-sky experiments}
at OHP using the PAPYRUS instrument \citep{chambouleyron2022first}, scheduled for mid 2024.

As prospective work, we can mention that the results obtained at $D/r_0=150$ are very encouraging to scale and extend our research towards ELTs--meaning orders-of-magnitude more Zernike modes and larger $D/r_0$ turbulence ranges--which can also benefit from the design and use of modern optical preconditioners \citep{Vera2023} to improve even further the dynamic range of the non-modulated PyWFS. Moreover, we may extend our work to detect differential piston modes between the ELT segments (petal modes) using the non-modulated pyramid \citep{levraud2022adapting}, while also adapting the GCViT for improving the \textcolor{black}{wavefront estimation performance} of an even more sensitive WFS such as the Zernike WFS \citep{cisse2022phase}.

\begin{acknowledgements}
The authors gratefully acknowledge the financial support provided by Agencia Nacional de Investigacion y Desarrollo (ANID) ECOS200010, STIC2020004, ANILLOS ATE220022, BECA DOCTORADO NACIONAL 21231967; Fondos de Desarrollo de la Astronomía Nacional (QUIMAL220006, ALMA200008); Fondo Nacional de Desarrollo Científico y Tecnológico (FONDECYT) (EXPLORACION 13220234, Postdoctorado 3220561); French National Research Agency (ANR) \emph{WOLF (ANR-18-CE31-0018)}, \emph{APPLY (ANR-19-CE31-0011)}, \emph{LabEx FOCUS (ANR-11-LABX-0013)}; Programme Investissement Avenir \emph{F-CELT (ANR-21-ESRE-0008)}, \emph{Action Spécifique Haute Résolution Angulaire (ASHRA)} of CNRS/INSU co-funded by CNES, \emph{ORP-H2020} Framework Programme of the European Commission’s (Grant number \emph{101004719}), Région Sud and the french government under the \emph{France 2030 investment plan}, as part of the \emph{Initiative d'Excellence d'Aix-Marseille Université A*MIDEX, program number AMX-22-RE-AB-151}. 
\end{acknowledgements}

\bibliographystyle{aa} 
\bibliography{references} 
    

\end{document}